\documentclass[a4paper,11pt]{article}
\usepackage{graphicx}
\usepackage{epsfig}
  
\newcommand{\Z}{{\sf Z\hspace*{-0.9ex}%
\rule{0.15ex}{1.5ex}\hspace*{0.9ex}}}
\newcommand{\N}{{\sf N\hspace*{-0.9ex}%
\rule{0.15ex}{1.5ex}\hspace*{0.9ex}}}

\title{A fractional generalization of the classical lattice dynamics approach}

\author{ {\sl T.M. Michelitsch$^{1}$\footnote{Corresponding author, e-mail~: michel@lmm.jussieu.fr },
B.A. Collet$^{1}$}, A.P. Riascos$^2$ \\ \\ {\sl A. F. Nowakowski$^3$, F.C.G.A. Nicolleau$^3$}
\\ \\
$^1$ Sorbonne Universit\'es \\ Universit\'e Pierre et Marie Curie (Paris 6) \\ Institut Jean le Rond d'Alembert, CNRS UMR 7190 \\
4 place Jussieu, 75252 Paris cedex 05, France
\\ \\
$^2$Instituto de F\'{i}sica, Universidad Nacional Aut\'{o}noma de M\'{e}xico\\  Apartado Postal 20-364, 01000 M\'{e}xico, D.F., M\'{e}xico
\\ \\
$^3$ Sheffield Fluid Mechanics Group\\
Department of Mechanical Engineering\\
University of Sheffield\\
Mappin Street, Sheffield S1 3JD,
United Kingdom
\\ \\ {\it  Chaos, Solitons \& Fractals, vol. 92, p. 43‑50, nov. 2016}\\ \\ {\sf We dedicate this paper to the memory of our esteemed friend and colleague G\'erard A. Maugin.}
}
\begin{document}

\maketitle

\newpage

\begin{abstract}
We develop physically admissible lattice models in the harmonic approximation which define by Hamilton's variational principle fractional Laplacian matrices of the forms of
power law matrix functions on the $n$-dimensional periodic and infinite lattice in $n=1,2,3,..$ dimensions. The present model which is based on
Hamilton's variational principle is confined to conservative non-dissipative isolated systems. The present approach yields the discrete analogue of
the continuous space fractional Laplacian kernel. As continuous fractional calculus generalizes differential operators such as 
the Laplacian to non-integer powers of Laplacian operators,
the fractional lattice approach developed in this paper generalized difference operators such as second difference operators to their fractional (non-integer) powers.
Whereas differential operators and difference operators constitute local operations, their fractional generalizations introduce nonlocal long-range features.
This is true for discrete and continuous fractional operators.
The nonlocality property of the lattice fractional Laplacian matrix allows to describe numerous anomalous transport phenomena such as anomalous fractional diffusion and random walks on lattices.
We deduce explicit results for the fractional Laplacian matrix in 1D for finite periodic and infinite linear chains and their Riesz fractional derivative continuum limit kernels.

The fractional lattice Laplacian matrix contains for $\alpha=2$ the classical local lattice approach
with well known continuum limit of classic local standard elasticity, and for other integer powers to gradient elasticity.
We also present a generalization of the fractional Laplacian matrix to n-dimensional cubic periodic (nD tori) and infinite lattices.
We show that in the continuum limit the fractional Laplacian matrix yields the well-known kernel of the Riesz fractional Laplacian
derivative being the kernel of the fractional power of Laplacian operator. In this way we demonstrate the interlink of the fractional
lattice approach with existing continuous fractional calculus. The developed approach appears to be useful to analyze fractional
random walks on lattices as well as fractional wave propagation phenomena in lattices.

\end{abstract}

\section{Introduction}

There are various phenomena in nature including complex, chaotic, turbulent, critical, fractal and anomalous transport phenomena having
erratic trajectories with often non-differentiable characteristics. Such `anomalous' phenomena as a rule cannot be described by standard
approaches involving integer order partial differential equations.
However, it has been shown
that they often can be described by non-integer order, i.e. fractional differential equations \cite{metzler,metzler2014}.

There are many definitions for fractional derivatives and integrals
(Riemann, Liouville, Caputo, Gr\"unwald-Letnikow, Marchaud, Weyl, Riesz, Feller, and others), see e.g. \cite{hilfer-2008,metzler,samko,samko2003,podlubny} and the references therein.
This diversity of definitions is due to the fact that fractional operators take different kernel representations in different function spaces which is a consequence of the nonlocal character of fractional kernels.

Whereas fractional operators
are well known in the continuous space and obtained as power law convolutional kernels, the fractional calculus on discrete networks and lattices is more involved
and much less developed.

An approach to define fractional differential operators on lattices
was suggested in the papers \cite{taraII,taraIII,tara3}. In \cite{tara1} was suggested the fractional calculus which includes fractional difference operators
and fractional integro-differential operators on lattices where exact results have been presented in these works (see also \cite{tara4}).
In contrast to that approach, the goal of the present work is to introduce {\it fractional centered difference operators} on the lattice
appearing as a natural fractional
generalization of Born von Karman's centered symmetric second order difference operator. In the same time all good properties of the classical Born von Karman lattice approach
such as translational symmetry and elastic stability is conserved by the fractional lattice model to be developed in the present paper. In this way the present
approach opens the door towards a generalization of the classical lattice approaches \cite{montroll}.

In the context of Markovian processes on networks, the concept of `fractional diffusion on undirected networks' generalizing the `normal random walk'
was recently introduced
by Riascos and Mateos
\cite{riascos12,riascos-fracdyn,riascos-fracdiff,riascos15}. Such random motions on lattices are defined by diffusion equations where instead of discrete Laplacian matrices
defined by second order difference operators their fractional generalizations come into play. In these works it has been demonstrated that fractional generalizations
of lattice models have a huge interdisciplinary potential as they are able to describe phenomena which account for nonlocal interactions
including the emergence of L\'evy flights on lattices \cite{riascos-fracdyn,riascos-fracdiff}.

Fractional generalizations second order difference operators introduce nonlocal long-range features. In contrast to fractional orders, the standard finite differences of integer orders
describe local nearest neighbor interactions. The nonlocality increases with the integer order of the power.
The difference operators of integer orders
at the lattice level correspond to spatial forms of non-local lattice sums
and the relations of differences and differential
operators defined by infinite series
are analyzed in recent articles \cite{tara2,tara3}, and further a fractional generalization of lattice derivative approach was introduced in \cite{tara4}.

Beside the applications on diffusion problems on the lattice, the importance of fractional lattice models appears also for a
description of fractional lattice vibrational phenomena, a generalization of crystal lattice dynamics. Some initial steps towards such a fractional
generalization generalization of nonlinear classical lattice dynamics has been introduced by
Laskin and Zaslavsky \cite{laskin2006} and Tarasov and Zaslavsky \cite{tara5}. In a lattice dynamics model which defines by Hamilton's variational principle the `Laplacian matrix'
which contains all constitutive
information of the harmonic interparticle interactions, it is therefore desirable to develop a
`fractional generalization' of the existing lattice dynamics approach. To this end in the present paper we utilize the methodology of characteristic matrix function
which was introduced recently \cite{michel-collet}.

The interdisciplinary fractional lattice approach developed in the present paper allows to describe all types of physical fields in the
harmonic approximation defined on lattices having a continuum limit governed by a non-classical, `anomalous' power-law nonlocality in the form of Riesz fractional
derivative kernels.
Physical applications include especially problems of fractionally generalized lattice dynamics: lattice vibrations, elastic wave propagation, nonlocal elasticity,
nonlocal electrodynamics, fractional quantum mechanics on lattices and networks, and last but
not least above mentioned anomalous transport and diffusion phenomena with `fractional random walk models' and emergence of L\'evy flights.
The entire set of problems of classical lattice dynamics in the harmonic approximation \cite{BornHuang}
can be fractionally generalized by the present approach. This generalization contains only one characteristic parameter, namely the power-law index $\alpha$. Newly emerging
phenomena of this `fractionalization' are to be expected.
The classical (Born von Karman) theories are contained in the approach at the value $\alpha=2$. Since the present fractional lattice
 approach is well defined by Hamilton's variational principle, it provides a physically admissible nonlocal generalization
 of any classical harmonic lattice model.

The present paper is organized as follows. In the first part of the paper
we deduce from ``fractional harmonic lattice potentials" on the cyclically closed linear chain a discrete fractional Laplacian matrix.
We do so by applying our recent approach to generate
nonlocal lattice models by matrix functions where the generator operator is the discrete centered Born von
Karman Laplacian \cite{michel-collet}. First we obtain the discrete fractional Laplacian
in explicit form for the infinite chain for particle numbers $N\rightarrow \infty$,
being in accordance with the fractional centered difference models of Ortiguiera \cite{riesz2}
and Zoia et al. \cite{Zoia2007}.
Utilizing the discrete infinite chain fractional Laplacian matrix we construct an explicit representation for the {\it fractional Laplacian matrix on the $N$-periodic finite 1D lattice} where the particle number
$N$ can be arbitrary not necessarily large.
Then we analyse continuum limits of the discrete fractional model:
The infinite space continuum limit of the fractional Laplacian matrix yields the well known infinite space kernel
of the standard fractional Laplacian. The periodic string continuum
limit yields an explicit representation for the kernel of the fractional Laplacian (Riesz fractional derivative) which fulfills periodic boundary conditions and is defined on the finite $L$-periodic string.

In the second part of the paper we suggest an extension of the fractional approach on nD periodic and infinite lattices. We deduce an integral representation for fractional Laplacian on the infinite nD lattice and proof that
as asymptotic representation the well known Riesz fractional derivative of the nD infinite space is emerging.
More detailed derivations of some of the results of the present paper can be found in recent articles
\cite{michelJphysA,michelchaos}. All these results are fully equivalent and can also be deduced by employing the more general
approach of Riascos and Mateos for fractional diffusion problems on networks \cite{riascos-fracdyn,riascos-fracdiff}, and see also the references therein.

\section{Fractional Laplacian matrix on cyclically closed chains}

We consider first a periodic, cyclically closed linear chain (1D periodic lattice or ring) with equidistant lattice points $p=0,..,N-1$ consisting of $N$
identical particles with particle mass $\mu$.
Each mass point $p$ has equilibrium position at $0\leq x_p=ph < L=Nh $ where $L$ denotes the length of the chain and $h$ the interparticle distance (lattice constant).
Further we impose periodicity (cyclic closure of the chain). For convenience of our demonstration we introduce the unitary shift operator $D$ defined by $ Du_p= u_{p+1}$ and its adjoint $D^{\dagger}= D^{-1}$ with $D^{\dagger}u_p = u_{p-1}$.
We employ periodic boundary conditions (cyclic closure of the chain) $u_p=u_{p+sN}$ ($s\in {\bf Z}$) and equivalently, cyclic index convention $p \rightarrow p \,\,\, mod\, (N)  \in \{0,1,..,N-1\}$.
Any elastic potential in the {\it harmonic approximation} defined on the 1D periodic lattice can be written in the representation \cite{michel-collet}
\begin{equation}
\label{compfo}
V_f = \frac{\mu}{2}\sum_{p=0}^{N-1}u_p^*f(2{\hat 1}-D-D^{\dagger})u_p = -\frac{1}{2} \sum_{p=0}^{N-1}\sum_{q=0}^{N-1} u_q^*\Delta_f(|p-q|)u_p ,
\end{equation}
where $\Delta_f(|p-q|) = -\mu f_{|p-q|}$ indicates the (negative-semidefinite) Laplacian $N\times N$-matrix, ${\hat 1}$ the identity matrix, and $f$ we refer to as the characteristic function:
Physically admissible, elastically stable and translational invariant positive elastic potentials require for the 1D periodic lattice (cyclic ring) that the
characteristic function $f$ which is defined as a scalar function to have the following properties $0<f(\lambda) < \infty$ for $0<\lambda \leq 4$ (elastic stability) and $f(\lambda=0)=0$ (translational invariance, zero elastic energy for uniform translations of the lattice). For the approach to be developed we propose the characteristic function to assume power law
form
\begin{equation}
\label{powerlaw}
f^{(\alpha)}(\lambda) = \Omega_{\alpha}^2\lambda^{\frac{\alpha}{2}} ,\hspace{2cm} \alpha >0 ,
\end{equation}
which fulfills for $\alpha >0$ and $\Omega_{\alpha}^2 >0$ the above required good properties for the characteristic function. $\Omega_{\alpha}$ denotes a dimensional constant of physical dimension $sec^{-1}$.
Note that $2{\hat 1}-D-D^{\dagger}$
is the central symmetric second difference operator which is defined by $(2{\hat 1}-D-D^{\dagger})u_p=2u_p-u_{p+1}-u_{p-1}$. The matrix function $f(2{\hat 1}-D-D^{\dagger})$ is in general a  self-adjoint (symmetric)
positive semidefinite $N\times N$-matrix function of the simple $N\times N$ generator matrix $[2{\hat 1}-D-D^{\dagger}]_{pq}=2\delta_{pq}-\delta_{p+1,q}-\delta_{p-1,q}$. It can be easily seen that $f(2{\hat 1}-D-D^{\dagger})$
has T\"oplitz structure, i.e. its additional symmetry consists in the form
$f_{pq}=f_{qp}=f_{|p-q|}$, $p,q=0,..N-1$ giving the fractional generalization of the Born von Karman centered difference operator. The fractional elastic potential has then with (\ref{compfo}) the representation
\begin{equation}
\label{Valpha}
V_{\alpha} = \frac{\mu\Omega_{\alpha}^2}{2} \sum_{p=0}^{N-1}u_p^*(2{\hat 1}-D-D^{\dagger})^{\frac{\alpha}{2}}u_p =
\frac{\mu}{2} \sum_{p=0}^{N-1}\sum_{q=0}^{N-1}u_q^*f^{(\alpha)}_{|p-q|}u_p ,
\end{equation}
with the matrix elements $f^{(\alpha)}_{|p-q|} =\Omega_{\alpha}^2[(2\,{\hat 1}-D-D^{\dagger})^{\frac{\alpha}{2}}]_{|p-q|}$
of the {\it fractional characteristic matrix function}. In full analogy to the negative semidefinite continuous Laplacian (second derivative operator) we define here the fractional Laplacian matrix as the negative semidefinite matrix defined through Hamilton's variational principle\footnote{The sign convention differs in many references, so e.g. in \cite{riascos-fracdyn,riascos-fracdiff} the fractional Laplacian matrix is defined positive semidefinite corresponding to the definition of the characteristic fractional operator $(2{\hat 1}-D-D^{\dagger})^{\frac{\alpha}{2}}$. }
\begin{equation}
\label{fractlap}
\Delta_{\alpha} u_p = -\frac{\partial}{\partial u_p}V_{\alpha} , \hspace{2cm}
(\Delta_{\alpha})_{|q-p|} = -\mu\Omega_{\alpha}^2 [(2{\hat 1}-D-D^{\dagger})^{\frac{\alpha}{2}}]_{|p-q|} = -\mu f^{(\alpha)}_{|p-q|}  .
\end{equation}
We emphasize again that the fractional Laplacian matrix $\Delta_{\alpha}=\Delta_{\alpha}^{\dagger}$ is self-adjoint having a symmetric T\"oplitz structure matrix representation
$(\Delta_{\alpha})_{pq}=(\Delta_{\alpha})_{qp}=(\Delta_{\alpha})_{|q-p|}=(\Delta_{\alpha}^{\dagger})_{pq}$.
This property appears naturally as the shift operators $D$ and their adjoint
$D^{\dagger}$ are symmetrically contained. This is true for all admissible $\alpha >0$ and includes $\alpha=2$ for
the conventional central symmetric second difference operator.
In this relation we have introduced the positive-semidefinite fractional characteristic matrix $f^{(\alpha)}_{|p-q|}$ which is in our convention up to the prefactor $-\mu$
identical with the fractional Laplacian matrix (which we define as negative-semidefinite being the fractional analogue of $\frac{d^2}{dx^2}$.
To determine the fractional Laplacian matrix it is useful to consider the spectral representation
\begin{equation}
 \label{spectralrepfrac}
 f^{(\alpha)}_{|p-q|} =\frac{\Omega_{\alpha}^2}{N}\sum_{\ell=0}^{N-1}e^{i\kappa_{\ell}(p-q)}\left(4\sin^2{\frac{\kappa_{\ell}}{2}}\right)^{\frac{\alpha}{2}} ,\hspace{0.5cm}
 \kappa_{\ell}= \frac{2\pi}{N}\ell
\end{equation}
where we take advantage of the $N$-periodicity of the chain, i.e. the eigenvectors of the fractional Laplacian matrix are ortho-normal Bloch-vectors having the components
$\frac{e^{i\kappa_{\ell}p}}{\sqrt{N}}$.
For the infinite chain in the limit $N\rightarrow \infty$ the matrix elements of the fractional Laplacian matrix (\ref{fractlap})
can be evaluated explicitly.
\begin{equation}
\label{fractlattice}
\begin{array}{l}
\displaystyle f^{(\alpha)}_{|p-q|}=\Omega_{\alpha}^2(2-D-D^{\dagger})^{\frac{\alpha}{2}}_{|p-q|} ,\hspace{1cm} p,q \in {\bf \Z}_0  ,\\  \\
\displaystyle f^{(\alpha)}_{|p|} ,
=\frac{\Omega_{\alpha}^2}{2\pi}\int_{-\pi}^{\pi}e^{i\kappa p}\left(4\sin^2{\frac{\kappa}{2}}\right)^{\frac{\alpha}{2}} {\rm d}\kappa ,\hspace{1cm} p \in {\bf \Z}_0.
\end{array}
\end{equation}
This expression can be obtained in explicit form \cite{michelJphysA,michelchaos,riascos-fracdyn,Zoia2007}
\begin{equation}
\label{matrixelei}
f^{(\alpha)}(|p|) = \Omega_{\alpha}^2\,\frac{\alpha!}{\frac{\alpha}{2}!(\frac{\alpha}{2}+|p|)!}(-1)^p\prod_{s=0}^{|p|-1}(\frac{\alpha}{2}-s) =
\Omega_{\alpha}^2 \,(-1)^p\, \frac{\alpha!}{(\frac{\alpha}{2}-p)!(\frac{\alpha}{2}+p)!} ,
\end{equation}
where we introduced the generalized factorial function $\beta ! = \Gamma(\beta+1)$. In view of (\ref{matrixelei})
we observe that for noninteger $\frac{\alpha}{2}$ any matrix element $f^{(\alpha)}(|p-q|) \neq 0$ is non-vanishing
indicating the nonlocality of the harmonic fractional interparticle interaction (\ref{fractlap}).
For $\frac{\alpha}{2} = m \in {\bf N}$ the matrix elements (\ref{matrixelei}) take the values of the standard binomial coefficients.
(\ref{fractlattice})$_2$ can be read as the Fourier coefficients of the infinite Fourier series
\begin{equation}
\label{Fouser}
\begin{array}{l}
\displaystyle \omega_{\alpha}^2(\kappa) = \Omega_{\alpha}^2\left(4\sin^2{\frac{\kappa}{2}}\right)^{\frac{\alpha}{2}} =
\displaystyle \Omega_{\alpha}^2\left(2-e^{i\kappa }-e^{-i\kappa }\right)^{\frac{\alpha}{2}} = \sum_{p=-\infty}^{\infty}f^{(\alpha)}_{|p|}e^{ip \kappa} \nonumber \\ \displaystyle \nonumber \\
 \hspace{1cm}\displaystyle  = f^{(\alpha)}_{0}+2\sum_{p=1}^{\infty}f^{(\alpha)}_{|p|}\cos{(p\kappa)}
 \end{array}
\end{equation}
representing the fractional dispersion relation.
Putting $\kappa=0$ in the dispersion relation, we can verify directly that the fractional Laplacian matrix conserves translational symmetry which is expressed by
\begin{equation}
\label{translationalsymmetry}
\sum_{p=-\infty}^{\infty}f^{(\alpha)}_{|p|} = 0
\end{equation}
This equation can also be read as $(2-D-D^{\dagger})^{\frac{\alpha}{2}} 1 =0$, i.e. the fractional centered difference operator applied to a constant is vanishing.
This property appears as the fractional generalization of the same property of second order centered differences when $\alpha=2$.
We further observe in view of (\ref{Fouser}) the positive semi-definiteness of the fractional characteristic matrix $f^{(\alpha)}_{|p-q|}$ where positiveness of
(\ref{Fouser}) for $0 < \kappa < 2\pi $ indicates elastic stability of the chain.

The fractional dispersion relation (\ref{Fouser}) leads to the remarkable
relation which holds {\it only} for complex numbers on the unit circle $z=e^{i\kappa }$, namely
\begin{equation}
\label{Fouserb}
\left(2-z-\frac{1}{z}\right)^{\frac{\alpha}{2}} = \sum_{p=-\infty}^{\infty}(-1)^p\, \frac{\alpha!}{(\frac{\alpha}{2}-p)!(\frac{\alpha}{2}+p)!}z^{p} ,
\hspace{2cm} |z|=1 .
\end{equation}
This Laurent series converges nowhere except on the unit circle $|z|=1$.
For instance the zero eigenvalue $\omega_{\alpha}^2(\kappa=0)=0$ which corresponds to translational invariance (zero elastic energy for uniform translations)
is obtained by putting $z=1$ in (\ref{Fouserb}).
For integer $\frac{\alpha}{2}=m \in {\bf N}$ (\ref{matrixelei}) takes the form of the standard binomial coefficients
and the series (\ref{Fouser}), (\ref{Fouserb}) then take the representations of standard binomial series of
$\left(2-z-\frac{1}{z}\right)^{\frac{\alpha}{2}}=(-1)^m(\sqrt{z}-\frac{1}{\sqrt{z}})^{2m}$ breaking at $|p|=m$ corresponding to zero values for the matrix elements for (\ref{matrixelei}) for $|p|>m$.
We further observe for noninteger $\frac{\alpha}{2} \notin {\bf N}$ the {\it power law asymptotics} for $|p|>>1$ which can be obtained by utilizing Stirling's asymptotic formula for the $\Gamma$-function \cite{michelJphysA,michelchaos}
\begin{equation}
\label{asymp}
f^{(\alpha)}_{|p|>>1} \rightarrow -\Omega_{\alpha}^2\,\,\frac{\alpha!}{\pi}\sin{(\frac{\alpha\pi}{2})} \,\, p^{-\alpha-1} .
\end{equation}

The asymptotic power law (scale free) characteristics of the fractional Laplacian matrix $\Delta_{pq} \sim |p-q|^{-\alpha-1}$ is the essential property which gives rise to many
`anomalous phenomena' such as in `fractional diffusion' problems on networks such as the emergence of L\'{e}vy flights \cite{riascos-fracdyn,riascos-fracdiff} (and references therein). The fractional continuum limit kernels are discussed in the subsequent section.
The expressions (\ref{fractlattice})-(\ref{asymp}) hold for the infinite 1D lattice corresponding to $N\rightarrow\infty$.
As everything in nature is limited we shall consider now the fractional Laplacian matrix for a finite periodic lattice
where the particle number $N$ is arbitrary and not necessarily large.
\newline\newline
{\bf 1D finite periodic lattice - ring}
\newline\newline
It is only a small step to construct the finite lattice Laplacian matrix in terms of infinite lattice Laplacian matrix.
We can perform this step by the following consideration: Let $-\mu f^{(\infty)}_{|p-q|}$ the Laplacian matrix of the infinite lattice, and $\omega^2(\kappa)$ the continuous dispersion relation of the infinite lattice matrix $f_{|p-q|}$ obeying the eigenvalue relation

\begin{equation}
\label{eigval}
\sum_{q=-\infty}^{\infty}f^{(\infty)}_{|p-q|}e^{iq\kappa} = \omega^2(\kappa)e^{ip\kappa} ,\hspace{2cm} 0\leq \kappa < 2\pi .
\end{equation}
This relation holds identically in the entire principal interval $0\leq \kappa < 2\pi$ and is $2\pi$-periodic in the $\kappa$-space. Let us now choose
$\kappa=\kappa_{\ell}=\frac{2\pi}{N}\ell$ with $\ell=0,..,N-1$ being the Bloch wave number of the
{\it finite} periodic lattice of $N$ lattice points where $N$ is not necessarily large. Since the Bloch wave numbers of the chain are discrete points within the interval $0\leq \kappa_{\ell}<2\pi$, then relation (\ref{eigval})
holds as well for these $N$ $\kappa$-points, namely
\cite{michelJphysA,michelchaos}\footnote{where $p=0$ in (\ref{eigval}) has been put to zero.}
\begin{equation}
\label{eigvalblock}
\begin{array}{l}
\displaystyle \sum_{p=-\infty}^{\infty}f^{(\infty)}_{|q|}e^{iq\kappa_{\ell}} =
\omega^2(\kappa_{\ell}) ,\hspace{2cm} 0\leq \kappa_{\ell}=\frac{2\pi}{N}\ell < 2\pi , \nonumber \\ \nonumber\\
\displaystyle \sum_{p=0}^{N-1}\sum_{s=-\infty}^{\infty}f^{(\infty)}_{|p+sN|} e^{i(p+sN)\kappa_{\ell}} = \sum_{p=0}^{N-1}e^{ip\kappa_{\ell}}\sum_{s=-\infty}^{\infty}f^{(\infty)}_{|p+sN|} =
\sum_{p=0}^{N-1}e^{ip\kappa_{\ell}} f^{finite}_{|p|} =  \omega^2(\kappa_{\ell}) .
\end{array}
\end{equation}
In the second relation the $N$-periodicity of the finite lattice Bloch eigenvector $e^{i(p+sN)\kappa_{\ell}}= e^{ip\kappa_{\ell}}$  has been used.
The last relation can be read as the eigenvalue relation for the $N$-periodic lattice matrix of T\"oplitz structure
\begin{equation}
\label{identiperiodic}
f^{finite}_{|p-q|} =
\sum_{s=-\infty}^{\infty}f^{(\infty)}_{|p-q+sN|} = f^{(\infty)}_{|p-q|}+ \sum_{s=1}^{\infty}(f^{(\infty)}_{|p-q+sN|}+f^{(\infty)}_{|p-q-sN|}) .
\end{equation}
It follows that in the limiting case $N\rightarrow\infty$ the finite lattice matrix (\ref{identiperiodic}) recovers the infinite lattice matrix $f^{finite} \rightarrow f^{(\infty)}$. From (\ref{identiperiodic}) we read of for the fractional lattice Laplacian of the finite periodic 1D lattice

\begin{equation}
\label{finitefraclap}
\Delta_{\alpha,N}(|p|) = -\mu f^{(\alpha,finite)}_{|p|} ,\hspace{1cm} 0\leq p \leq N-1
\end{equation}
with
\begin{equation}
\label{finitecharmat}
\begin{array}{l}
f^{(\alpha,finite)}_{|p|} =  \Omega_{\alpha}^2\displaystyle \frac{(-1)^p\alpha!}{(\frac{\alpha}{2}-p)!(\frac{\alpha}{2}+p)!} + \Omega_{\alpha}^2 \sum_{s=1}^{\infty}(-1)^{p+Ns}\alpha !
\left(  \frac{1}{(\frac{\alpha}{2}-p-sN)!(\frac{\alpha}{2}+p+sN)!} \right. \\ \\
\left. \hspace{2cm} \displaystyle + \frac{1}{(\frac{\alpha}{2}-p+sN)!(\frac{\alpha}{2}+p-sN)!} \right) .
\end{array}
\end{equation}
We observe $N$-periodicity of (\ref{finitecharmat}) and furthermore the necessary property that in the limit of infinite chain $N\rightarrow \infty$, (\ref{finitecharmat}) recovers the infinite lattice expression
of eq. (\ref{matrixelei}).

\section{Fractional continuum limit kernels}

In this section we investigate the interlink between the lattice fractional approach introduced above and continuum fractional derivatives. To this end we introduce the following hypotheses which are to be observed when performing
continuum limits. Following \cite{michel-collet} we require in the continuum limit that extensive physical quantities, i.e. quantities
which scale with the length of the 1D system, such as the total mass $N\mu=M$ and the total elastic
energy of the chain remain finite when its length $L$ is kept finite\footnote{In the case of infinite string $L\rightarrow \infty$ we require the
mass per unit length and elastic energy per unit length to remain finite.}, i.e. neither vanish nor diverge. Let $L=Nh$ be the length of the chain and $h$ the lattice constant
(distance between two neighbor atoms or lattice points).

\noindent We can define two kinds of continuum limits: \newline\noindent  (i) The {\it periodic string continuum limit} where the length of the chain
$L=Nh$ is kept finite and $h\rightarrow 0$ (i.e. $N(h)= L h^{-1} \rightarrow \infty$). \newline\noindent (ii) The {\it infinite space continuum limit} where $h\rightarrow 0$,
however, the length of the chain tends to infinity $N(h)h=L(h) \rightarrow \infty$\footnote{which can be realized for instance by choosing by $N(h)\sim h^{-\delta}$ where $\delta > 1$.}. The kernels of the infinite space limit can be recovered from those of
the periodic string limit by letting $L\rightarrow \infty$.
From the finiteness of
the total mass of the chain, it follows that the particle mass $\mu=\frac{M}{N}=\frac{M}{L} h = \rho_0 h$ scales as $\sim h$.
 Then by employing expression (\ref{Valpha}) for the fractional elastic potential,
the total continuum limit elastic energy ${\tilde V}_{\alpha}$ can be defined by
\begin{equation}
\label{elasten}
{\tilde V}_{\alpha} =
\lim_{h\rightarrow 0+} V_{\alpha} =
\frac{\mu\Omega_{\alpha}^2}{2}\sum_{p=0}^{N-1} u^*(x_p)\left(-4\sinh^2{\frac{h}{2}\frac{d}{dx}}\right)^{\frac{\alpha}{2}}u(x_p) .
\end{equation}
Putting $D=e^{h\frac{d}{dx}}$ ($ph=x_p \rightarrow x$) and accounting for $2-D(h)-D(-h) =-4\sinh^2{\frac{h}{2}\frac{d}{dx}} \approx -h^2\frac{d^2}{dx^2} + O(h^4)$ we get
\begin{equation}
\label{fraclap}
\lim_{h\rightarrow 0} \left(-4\sinh^2{\frac{h}{2}\frac{d}{dx}}\right)^{\frac{\alpha}{2}} = h^{\alpha} (-\frac{d^2}{dx^2})^{\frac{\alpha}{2}} .
\end{equation}

The formal relation (\ref{fraclap}) shows that the continuum limit kernels to be deduced in explicit forms have the interpretation of the {\it Fractional Laplacian} or also in the literature referred to as {\it Riesz Fractional Derivative}.
To maintain finiteness of the elastic energy in the continuum limit $h\rightarrow 0$ the following scaling relations for the characteristic model constants, the mass $\mu$ and the frequency $\Omega_{\alpha}$ are required
\cite{michelJphysA,michelchaos}
\begin{equation}
\label{scaling}
\Omega_{\alpha}^2(h) =A_{\alpha} h^{-\alpha} , \hspace{1cm} \mu(h) = \rho_0 h ,\hspace{1cm} A_{\alpha}, \rho_0 >0
\end{equation}
where $\rho_0$ denotes the mass density with dimension $g \times cm^{-1}$ and $A_{\alpha}$ denotes a positive dimensional constant
of dimension $ sec^{-2}\times cm^{\alpha}$, where the new constants $\rho_0, A_{\alpha}$ are independent of $h$. Note that the dimensional
constant $A_{\alpha}$ is only defined up to a non-dimensional positive
scaling factor as its absolute value does not matter due to the scale-freeness of the power law.
We obtain then as continuum limit of the elastic energy by taking into account $\sum_{p=0}^{N-1}h G(x_p) \rightarrow \int_0^L G(x){\rm d}x$ and $h\rightarrow dx$,
$x_p\rightarrow x$,
\begin{equation}
\label{contilimielasten}
\begin{array}{l}
\displaystyle
{\tilde V}_{\alpha} = \lim_{h\rightarrow 0} \frac{\mu(h)}{2}
\sum_{q=0}^{N-1}\sum_{p=0}^{N-1}u_q^* f^{(\alpha)}_N(|p-q|)u_p \nonumber \\ \nonumber \\
\displaystyle
{\tilde V}_{\alpha} = \frac{\rho_0 A_{\alpha}}{2}
\int_0^Lu^*(x)\left(-\frac{d^2}{dx^2}\right)^{\frac{\alpha}{2}}u(x)\,{\rm d}x
=: -\frac{1}{2} \int_0^L\int_0^Lu^*(x'){\tilde \Delta}_{\alpha}(|x-x'|)u(x){\rm d}x{\rm d}x' .
\end{array}
\end{equation}
The continuum limit Laplacian kernel ${\tilde \Delta}_{\alpha}(|x-x'|)$ can then formally be represented by the distributional kernel
representation in the spirit of generalized functions \cite{gelfand}
\begin{equation}
\label{contilimlap}
{\tilde \Delta}_{\alpha,L}(|x-x'|) = -\rho_0A_{\alpha}\left(-\frac{d^2}{dx^2}\right)^{\frac{\alpha}{2}}\delta_L(x-x') .
\end{equation}

The last relation contains the distributional representation of the fractional Laplacian and is obtained for the infinite space limit (ii) in explicit form as (where $\Re(..)$ indicates the real part of $(..)$)
\cite{michelJphysA,michelchaos}
\begin{equation}
 \label{inffyla}
 {\cal K}_{\infty}^{(\alpha)}(x)= -\left(-\frac{d^2}{dx^2}\right)^{\frac{\alpha}{2}}\delta_L(x-x')=-\frac{\alpha !}{\pi} \lim_{\epsilon\rightarrow 0+} \Re
\frac{i^{\alpha+1}}{(x+i\epsilon)^{\alpha+1}} ,
\end{equation}
being defined `under the integral' in the distributional sense and yields for noninteger $\frac{\alpha}{2}\notin {\bf N}$ for $x\neq 0$ the well known Riesz fractional derivative kernel
of the infinite space ${\cal K}_{\infty}^{(\alpha)}(x)= \frac{\alpha ! \sin{(\frac{\alpha\pi}{2})}}{\pi}\frac{1}{|x|^{\alpha+1}}$ with a characteristic
$|x|^{-\alpha-1}$ power law nonlocality reflecting the asymptotic
power law behavior (\ref{asymp}) of (\ref{matrixelei}) for sufficiently large $|p|>>1$.

\subsection{(i) Periodic string continuum limit}
The continuum procedure of $L$-periodic string where $L$ is kept finite is then obtained as \cite{michelJphysA,michelchaos}\footnote{where $\Re(..)$ denotes the real part of a quantity $(..)$}

\begin{equation}
\label{perfrac}
\begin{array}{l}
\displaystyle -\left(-\frac{d^2}{dx^2}\right)^{\frac{\alpha}{2}}\delta_L(x) =
K_L^{(\alpha)}(|x|) = \frac{\alpha ! \sin{(\frac{\alpha\pi}{2})}}{\pi} \sum_{n=-\infty}^{\infty}\frac{1}{|x-nL|^{\alpha+1}}  ,\hspace{0.5cm} \xi=\frac{x}{L} ,\\  \\
\displaystyle \hspace{3cm}K_L^{(\alpha)}(|x|) = \frac{\alpha !\sin{(\frac{\alpha\pi}{2})}}{\pi L^{\alpha+1}}\left\{-\frac{1}{|\xi|^{\alpha+1}}+
{\tilde \zeta}(\alpha +1,\xi) + {\tilde \zeta}(\alpha +1,-\xi) \right\}   \\  \\
\displaystyle \hspace{3cm}K_L^{(\alpha)}(|x|) = -\frac{\alpha !}{\pi} \lim_{\epsilon\rightarrow 0+} \Re\left\{\sum_{n=-\infty}^{\infty}
\frac{i^{\alpha+1}}{(x-nL+i\epsilon)^{\alpha+1}}\right\}  \\ \\
\hspace{0.5cm}\displaystyle =\frac{\alpha !}{\pi L^{\alpha+1}} \lim_{\epsilon\rightarrow 0+}
\Re \left\{\ i^{\alpha+1} \left(  \frac{1}{(\xi+i\epsilon)^{\alpha+1}}
-\zeta(\alpha+1,\xi+i\epsilon)-\zeta(\alpha+1,-\xi+i\epsilon) \right)\right\} .
\end{array}
\end{equation}

This kernel can be conceived as the explicit representation of the fractional Laplacian (Riesz fractional derivative) on the $L$-periodic string.
The last relation is the distributional representation and is expressed by standard Hurwitz $\zeta$-functions denoted by $\zeta(..)$. The two variants of $\zeta$- functions which occur in above relation are defined by

\begin{equation}
\label{hurwitz}
{\tilde \zeta}(\beta,x)= \sum_{n=0}^{\infty}\frac{1}{|x+n|^{\beta}} ,\hspace{2cm} \zeta(\beta,x)=\sum_{n=0}^{\infty}
\frac{1}{(x+n)^{\beta}} ,\hspace{1.5cm} \Re\, (\beta) >1 , \hspace{0.5cm} \Re\ (x) >0 .
\end{equation}
We see for $\alpha>0$ and $x\neq 0$ that the series in (\ref{perfrac}) are absolutely convergent as good as the power function integral  $\int_1^{\infty}\xi^{-\alpha-1}{\rm d}\xi$. For integer powers $\frac{\alpha}{2} \in {\bf N}$ the distributional representations (\ref{perfrac})$_{3,4}$ take the (distributional) forms of the
(negative semi-definite) 1D integer power Laplacian operators, namely

\begin{equation}
\label{integerperiodic}
\begin{array}{l}
\displaystyle K_L^{(\alpha=2m)}(|x|)= (-1)^{m+1}\frac{d^{2m}}{dx^{2m}}
\sum_{n=-\infty}^{\infty} \lim_{\epsilon\rightarrow 0+}\frac{1}{\pi}\frac{\epsilon}{((x-nL)^2+\epsilon^2)} ,\hspace{1cm} \frac{\alpha}{2} = m\in {\bf N_0} ,\\ \\
\displaystyle \hspace{3cm} =
 (-1)^{m+1}\frac{d^{2m}}{dx^{2m}} \sum_{n=-\infty}^{\infty}\delta_{\infty}(x-nL) = -\left(-\frac{d^2}{dx^2}\right)^{\frac{\alpha}{2}=m}\delta_L(x) ,
\end{array}
\end{equation}
where $\delta_{\infty}(..)$ and $\delta_L$ indicate the Dirac's $\delta$-functions of the infinite and the $L$-periodic string, respectively.
We further observe in full correspondence to the discrete fractional Laplacian matrix, the necessary property that in the limit of an
infinite string $\displaystyle \lim_{L\rightarrow\infty}  K_L^{(\alpha)}(|x|) = {\cal K}_{\infty}^{(\alpha)}(x) $ (\ref{perfrac}) recovers the expression of the standard
1D infinite space fractional Laplacian kernel (\ref{inffyla})
known from the literature
(see for a further discussion \cite{michelJphysA,michelchaos} and references therein).

\section{Fractional Laplacian matrix on $n$-dimensional cubic lattices}

In this section we deduce the $nD$ counterpart of the fractional Laplacian matrix introduced above. With that approach the fundamentals of
`{\it fractional lattice dynamics}' can be deduced as a generalization of conventional lattice dynamics.

In this section our goal is to generalize the above 1D lattice approach to cubic periodic lattices in $n=1,2,3,..$
dimensions of the physical space where the 1D lattice case is contained. We assume the lattice contains $N=N_1..\times N_n$ lattice points, each covered by identical atoms with mass $\mu$.
Each mass point is characterized by $\vec{p}=(p_1,p_2,..,p_n)$ ($p_j=0,..N_j-1$) and $n=1,2,3,..$ denotes the dimension of the physical space embedding the lattice. In order to define the lattice
fractional Laplacian matrix, it is sufficient to consider
a {\it scalar} generalized displacement field $u_{\vec{p}}$ (one field degree of freedom) associated to each mass point $\vec{p}$ only. The physical nature of this scalar field can be any scalar field,
such as for instance a one degree of freedom displacement field, an
electric potential or, in a stochastic context a probability density function (pdf) or in a fractional quantum mechanics context a Schr\"odinger wave function. This demonstrates the interdisciplinary
character of the present fractional lattice approach.

The fractional Laplacian matrix for general networks was only recently and to our knowledge for the first time introduced by Riascos and Mateos \cite{riascos-fracdyn,riascos-fracdiff} in the framework of fractional diffusion analysis
on networks which include nD periodic lattices (nD tori) as special cases being subject of the present analysis. For cubic nD lattices the fractional Laplacian matrix can be written as
\cite{michelJphysA,michelchaos,riascos-fracdyn,riascos-fracdiff}

\begin{equation}
\label{fraalphn}
\Delta_{\alpha,n} = -\mu \Omega_{\alpha,n}^2L_n^{\frac{\alpha}{2}} ,\hspace{1cm} L_n^{\frac{\alpha}{2}} = \left(2n {\hat 1}-A_n \right)^{\frac{\alpha}{2}} ,
\hspace{0.5cm} \alpha > 0 ,
\end{equation}
where ${\hat 1}$ denotes the identity matrix, $n$ indicates the dimension of the physical space and $2n$ indicates the connectivity, i.e. the number of next neighbors of a lattice point in the nD cubic lattice.
In (\ref{fraalphn}) we introduced the adjacency matrix $A_n$ which has for the cubic lattice with next neighbor connections the form

\begin{equation}
\label{cubicadj}
A_n = \sum_{j=1}^n (D_j+D_j^{\dagger}) ,
\end{equation}
where then $D_j$ and $D_j^{\dagger}=D_j^{-1}$ denote the next neighbor shift operators in the $j=1,..,n$-directions defined by $D_ju_{p_1,..p_j,..,p_n}= \vec{u}_{p_1,..p_j+1,..,p_n}$
and $D_j^{\dagger}\vec{u}_{p_1,..p_j,..,p_n}= u_{p_1,..p_j-1,..,p_n}$, i.e.
$D_j$ shifts the field associated to lattice point $\vec{p}=(..,p_j,..$ to the field associated with the adjacent lattice point in the positive $j$-direction $(..,p_{j+1},..)$, and the inverse (adjoint) shift operator
$D_j^{\dagger}=D_j^{-1}$ to the adjacent lattice point in the negative $j$-direction $(..,p_{j-1},..)$. All matrices introduced in (\ref{fraalphn})
and (\ref{cubicadj})
are defined on the
nD lattice being $N \times N$ matrices ($N=N_1 \times ..\times N_n$).
As in the case of 1D lattice the so defined fractional Laplacian matrix (\ref{fraalphn}) describes for non-integer
powers $ \frac{\alpha}{2}$, $\notin {\bf N} $ nonlocal elastic interactions, whereas they are generated by the `local' next neighbor Born von Karman Laplacian which is in our definition up to a negative
dimension factor $-\mu\Omega_{2}$ equal to $L_n$. We therefore refer to $L_n$ as `generator matrix'. We emphasize that the sign convention of what
we call `(fractional) Laplacian matrix' varies in the literature
(e.g. by denoting the positive semidefinite matrix $L_n^{\frac{\alpha}{2}}$ as `fractional Laplacian matrix', this convention is chosen, e.g. in \cite{riascos-fracdyn,riascos-fracdiff}).
We have chosen to refer to as `fractional Laplacian matrix'
the negative-semidefinite matrix
$-\mu\Omega_{\alpha}^2L_n^{\frac{\alpha}{2}}$ to be in accordance with the negative definiteness of continuum limit fractional Laplacian (\ref{integerperiodic}) containing as a special case $\frac{\alpha}{2}=1$ the negative semidefinite
conventional Laplacian $\frac{d^2}{dx^2}\delta_L(x-x')$).
For a discussion of some general properties of the fractional Laplacian (\ref{fraalphn}) well defined on general networks including $nD$ lattices, we refer to \cite{riascos-fracdyn,riascos-fracdiff}.
In the periodic and infinite lattice the shift operators are unitary.
Assuming $N_j$-periodicity in each direction $j$, the fractional Laplacian matrix is defined by
the spectral properties of the $L_n$-matrix, namely by
\begin{equation}
\label{spectralfrac}
[L_n^{\frac{\alpha}{2}}]_{(\vec{p}-\vec{q})}=\frac{1}{N}
\sum_{\vec{\ell}} e^{i\vec{\kappa}_{\vec{\ell}}\cdot(\vec{p}-\vec{q})}\lambda_{\vec{\ell}}^{\frac{\alpha}{2}} ,
\hspace{0.5cm} \lambda_{\vec{\ell}} = \left(2n-2\sum_{j=1}^n\cos{(\kappa_{\ell_j})}\right) ,\,\alpha >0,
 \end{equation}
where we denoted $\sum_{\vec{\ell}}(..)=\sum_{\ell_1=0}^{N_1-1}(..)..\sum_{\ell_n=0}^{N_n-1}(..)$ and $\vec{\kappa}_{\vec{\ell}}=(\kappa_{\ell_1},..\kappa_{\ell_n}) $ denotes the Bloch wave vectors of the Brillouin zone where their components
can the values $\kappa_{\ell_j}=\frac{2\pi}{N_j}\ell_j$ ($\ell_j=0,..,N_j-1$).
It can be seen that (\ref{spectralfrac}) has T\"oplitz structure depending only on $|p_1-q_1|,..,|p_j-q_j|,..|p_n-q_n|$).
For the infinite lattice when all $N_j\rightarrow\infty$ in (\ref{spectralfrac}), the summation over the
reciprocal lattice points assumes asymptotically the form of an integral
$\frac{1}{N}\sum_{\vec{\ell}}g(\vec{\kappa}_{\ell}) \sim \frac{1}{(2\pi)^n}\int_{-\pi}^{\pi}..\int_{-\pi}^{\pi}{\rm d}\kappa_1..{\rm d}\kappa_n
g(\vec{\kappa})$, where the integration intervals $[-\pi,\pi]$ can be chosen instead of $[0,2\pi]$ for $2\pi$-periodic functions $g(\kappa_j)=g(\kappa_j+2\pi)$.
\vskip0.5cm
\begin{center}
\includegraphics[scale=0.32]{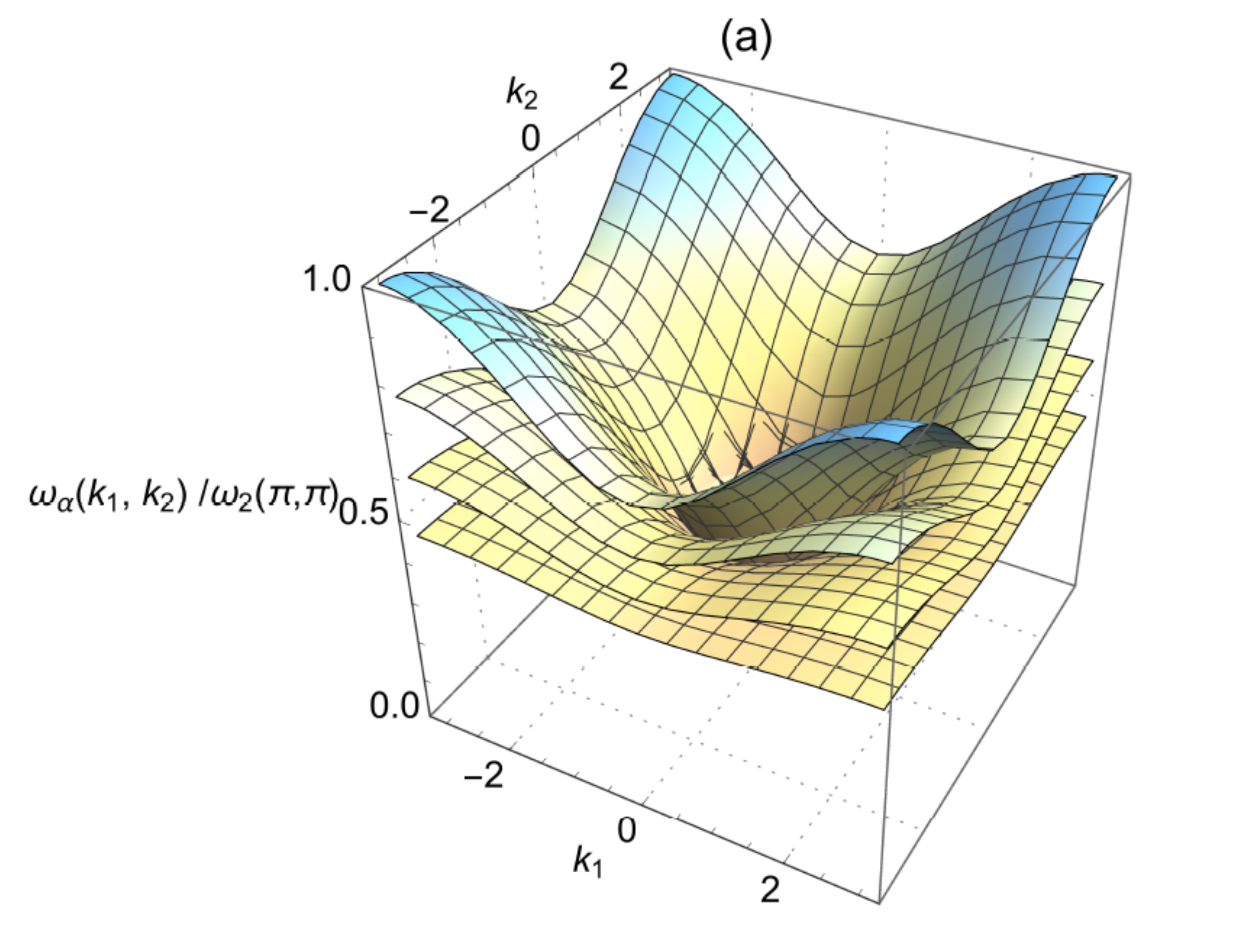}
\end{center}
\vskip-2cm
\begin{center}
\includegraphics[scale=0.32]{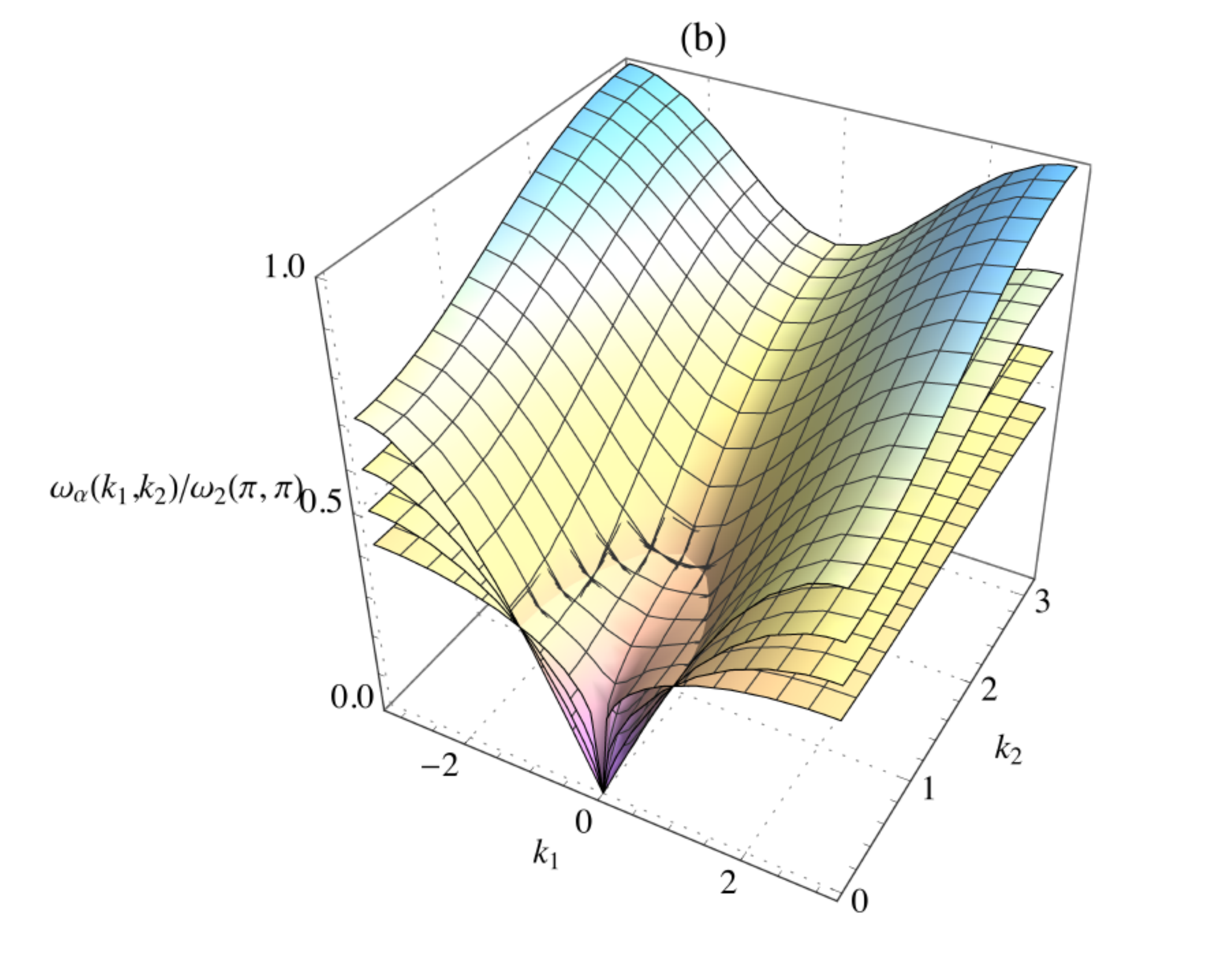}
\end{center}
\begin{center}
\includegraphics[scale=0.32]{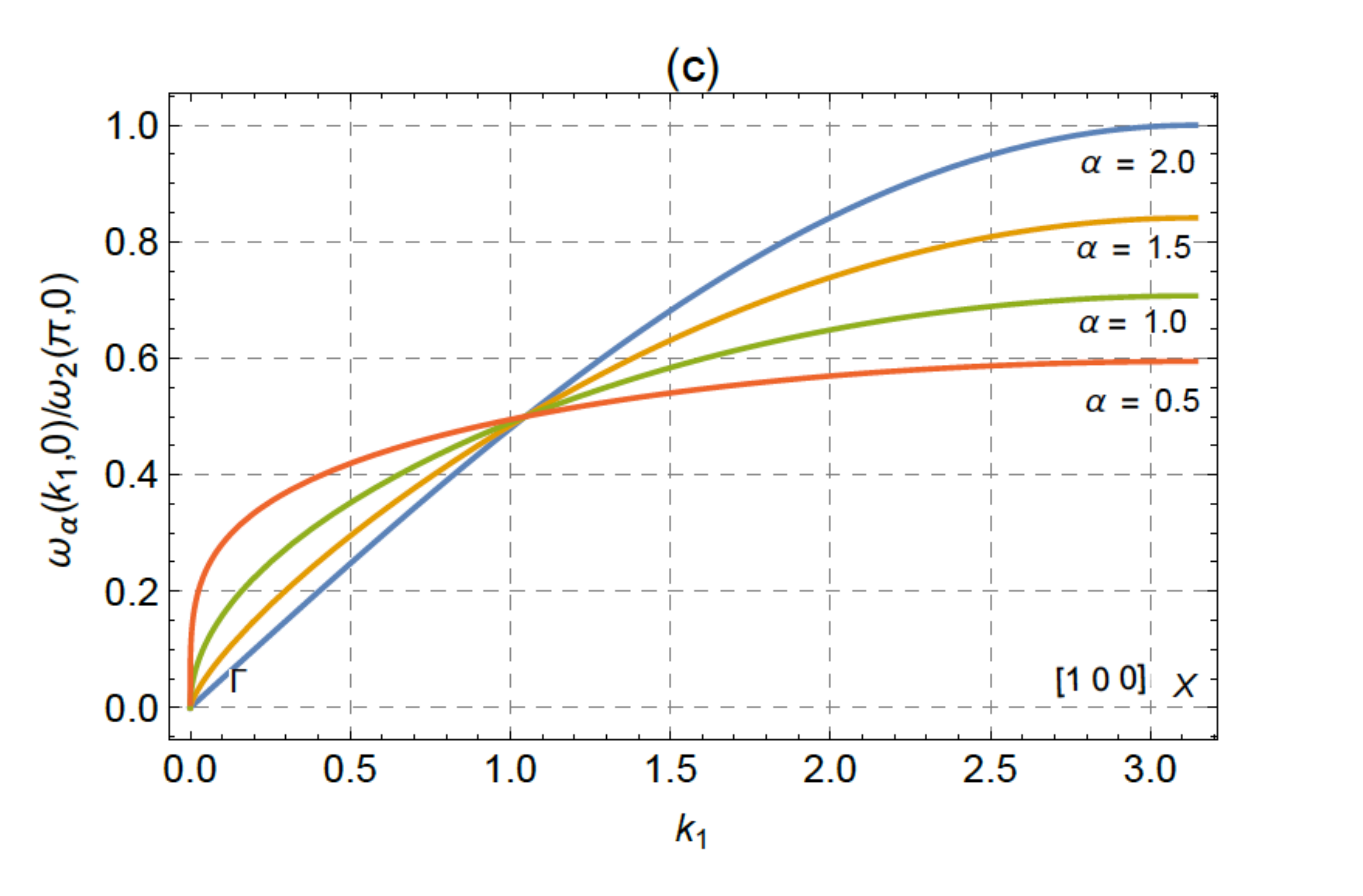}
\end{center}
\begin{center}
\includegraphics[scale=0.32]{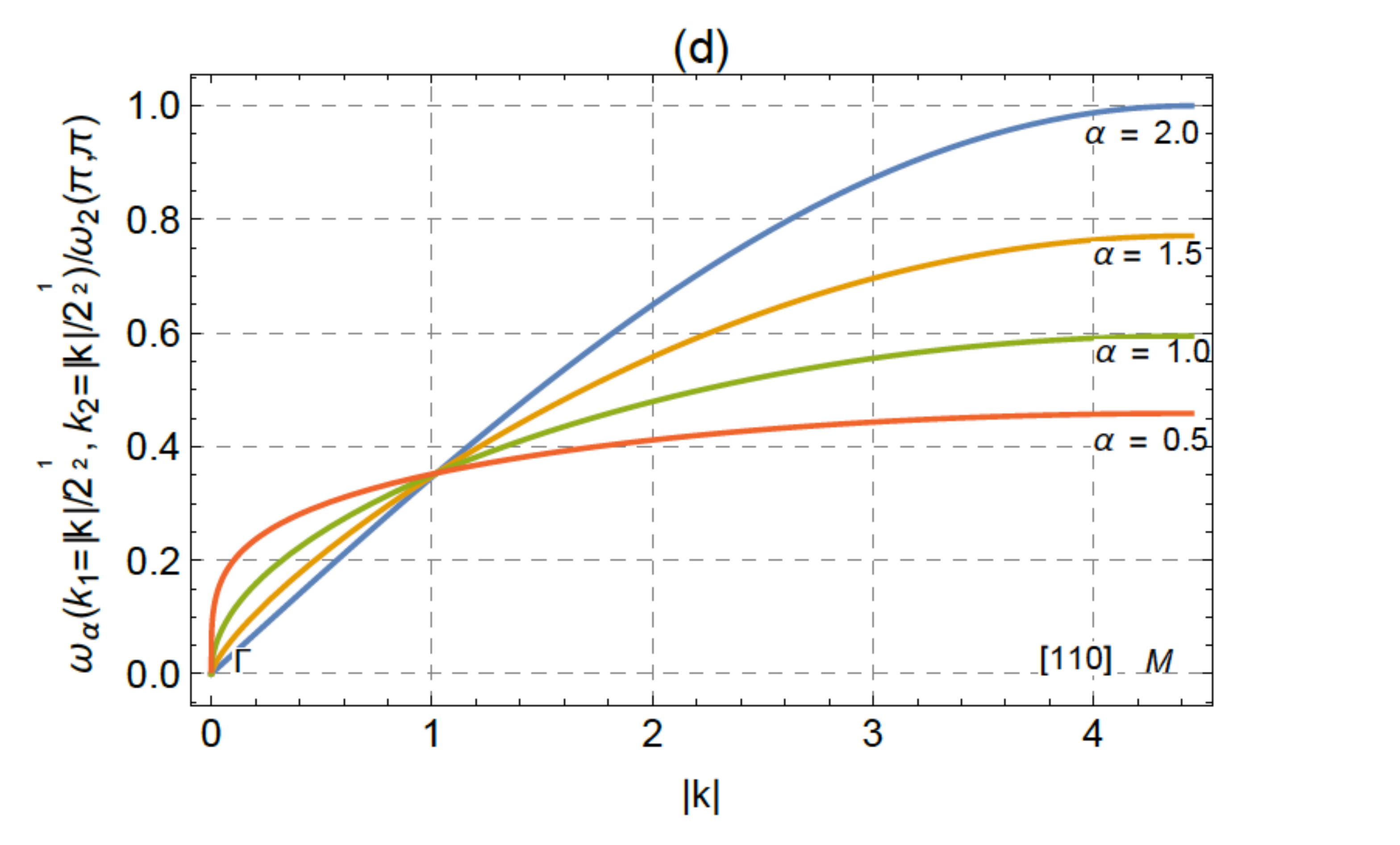}
\end{center}
{\it ${\bf{Fig. 1}}$ (a-b)
Show the dispersion surfaces $\omega_{\alpha}(\kappa_1,\kappa_2)/\omega_{2}(\pi,\pi)=
\lambda^{\frac{\alpha}{4}}(\kappa_1,\kappa_2)/2^{\frac{3}{2}} = 2^{\frac{\alpha-3}{2}} (sin^2(\kappa_1/2)+ sin^2(\kappa_2/2))^{\frac{\alpha}{4}} $ for the 2D cubic
lattice ($n=2$) of (\ref{spectralfrac}) for four values of $\alpha$, while (c-d) illustrate cross-sections of these dispersion sheets
with the planes (0 1 0) and (1 1 0), respectively.}
 \newline\newline
 For  $\alpha$ fixed, the circular frequency is given by $\omega_{\alpha}(\kappa_1,\kappa_2) = \lambda^{\frac{\alpha}{4}}$.
 The linear frequency spectra $(a, b, d)$, for $n=2$, are normalized by the maximum frequency
  $\omega_{\alpha=2}(\pi,\pi)= \lambda^{\frac{1}{2}}(\pi,\pi)= 2^{\frac{3}{2}} $ obtained for
  a wave vector located in $(001)$ plane. It will be noted that the sheets cut at dimensionless frequency
 $\omega_{\alpha}(\kappa_1,\kappa_2)/\omega_{\alpha=2}(\pi,\pi) \approx 0.351$
 and the dispersion relations of the classical next neighbor Born von Karman lattice are recovered (indicated by
 $\omega_{\alpha}(\kappa_1,\kappa_2)/\omega_{\alpha=2}(\pi,\pi) \rightarrow 1$ for $\alpha=2$ and $\kappa_{1,2}\rightarrow \pi$,
 $\omega_{\alpha}(\kappa_1,0)/\omega_{\alpha=2}(\pi,0) \rightarrow 1$ for $\alpha=2$ and $\kappa_1\rightarrow \pi$).
 When the value of $\alpha$ decreases, one observes
 in agreement with another work \cite{michelJphysA}, namely a decrease of the maximum dimensionless frequency in end of the first Brillouin zone.

The goal is now to deduce a more convenient integral representation of (\ref{spectralfrac}). To this end we utilize the following observation: Let in the following ${\cal L}$ be a positive
semidefinite\footnote{i.e. all eigenvalues $\lambda_{\ell}$
of this matrix are non-negative.} matrix and $\alpha >0$
 like (\ref{fraalphn}) with the spectral representation
\begin{equation}
 \label{spere}
{\cal L}=  \sum_{\vec{\ell}}\lambda_{\vec{\ell}} |\vec{\ell}><\vec{\ell}| ,\hspace{2cm} {\cal L}_{pq} = <p|{\cal L}q> ,
\end{equation}
where we have to put for the periodic nD lattice of (\ref{spectralfrac}) the Bloch-eigenvectors $<\vec{p}|\vec{\ell}> = N^{-\frac{1}{2}} e^{i\vec{\kappa}_{\vec{\ell}}\cdot\vec{p}}$.
Then it will be useful to define the matrix Dirac $\delta$-function by
\begin{equation}
 \label{matrixdelta}
 \delta({\cal L}-\tau{\hat 1}) = \sum_{\vec{\ell}} |\vec{\ell}><\vec{\ell}|\, \delta(\tau-\lambda_{\vec{\ell}}) ,
\end{equation}
where $\tau$ is a scalar parameter and ${\hat 1}$ the identity matrix and $\delta(\tau-\lambda_{\vec{\ell}})$
the conventional scalar Dirac $\delta$-function. Then with the matrix $\delta$-function defined in
(\ref{matrixdelta}) we can write
\begin{equation}
 \label{matrixpower}
{\cal L}^{\frac{\alpha}{2}} = \int_{-\infty}^{\infty}\delta(L-\tau{\hat 1})|\tau|^{\frac{\alpha}{2}}{\rm d}\tau
\end{equation}
and by utilizing $\delta(\tau-\lambda_{\vec{\ell}}) = \frac{1}{(2\pi)}\int_{-\infty}^{\infty}e^{ik(\tau-\lambda_{\vec{\ell}})}{\rm d}k$
together with the kernel $-{\cal D}_{\frac{\alpha}{2}}$ of the 1D fractional Laplacian (Riesz fractional derivative) of order $\frac{\alpha}{2}$ in its distributional form \cite{michel-et-al-14}
\begin{equation}
 \label{distrifoufracder}
 \begin{array}{l}
 {\cal D}_{\frac{\alpha}{2}}(k-\xi) = \displaystyle \left(-\frac{d^2}{dk^2}\right)^{\frac{\alpha}{4}}\delta(k-\xi) =:
\frac{1}{(2\pi)}\int_{-\infty}^{\infty}e^{i(k-\xi)\tau}|\tau|^{\frac{\alpha}{2}}{\rm d}\tau =  \\ \\
\displaystyle \lim_{\epsilon\rightarrow 0+}\frac{1}{\pi}\Re \int_0^{\infty}e^{-\tau(\epsilon-i(k-\xi))}|\tau|^{\frac{\alpha}{2}}{\rm d}\tau =
 \lim_{\epsilon\rightarrow 0+} \Re \frac{\Gamma(\frac{\alpha}{2}+1)}{\pi(\epsilon-i(k-\xi))^{\frac{\alpha}{2}}} .
\end{array}
\end{equation}
Then we can write for the matrix power function (\ref{matrixpower}) the representation
\begin{equation}
 \label{repre}
 {\cal L}^{\frac{\alpha}{2}} = \int_{-\infty}^{\infty} e^{ik{\cal L}}{\cal D}_{\frac{\alpha}{2}}(k) {\rm d}k ,
\end{equation}
where the exponential $e^{ik{\cal L}}$ of the matrix ${\cal L}=L_n$ can be determined more easily for the generator $L_n = \sum_{j=1}^n L_{(j)}$  ($L_{(j)}=2-D_j-D_j^{\dagger}$) being the sum of the 1D generator matrices of the $N_j$-periodic
1D lattices and having therefore the eigenvalues $\lambda(\ell_j)=2-2\cos{\kappa_{\ell_j}}$ and as a consequence having a Cartesian product space spanned by the periodic Bloch eigenvectors
$\frac{e^{i{\vec \kappa}_{\vec{\ell}}\cdot \vec{p}}}{\sqrt{N}} =\prod_{j=1}^n \frac{e^{ip_j\kappa_{\ell_j}}}{\sqrt{N_j}} $.
The matrix elements of the spectral representation of the exponential of $L_n$ can hence be written as
\begin{equation}
 \label{expolapl}
 [e^{i\xi L_n}]_{\vec{p}-\vec{q}} = \sum_{\vec{\ell}} \frac{e^{i{\vec \kappa}_{\vec{\ell}}\cdot (\vec{p}-\vec{q})}}{N}e^{i\xi \lambda_{\vec{\ell}}} =
 \prod_{j=1}^n \sum_{\ell_j=1}^{N_j-1} \frac{e^{i(p_j-q_j)\kappa_{\ell_j}}}{N_j}e^{i2k(1-\cos{\kappa_{\ell_j}})} .
\end{equation}

\noindent {\bf Infinite nD lattice}\\
In the limiting case of an infinite nD lattice when all $N_j\rightarrow\infty$ we can write by using $\frac{1}{N}\sum_{\vec{\ell}}g({\vec \kappa}_{\vec{\ell}}) \sim \frac{1}{(2\pi)^n}
\int_{-\pi}^{\pi}..\int_{-\pi}^{\pi}{\rm d}\kappa_1..{\rm d}\kappa_n g({\vec \kappa}) $ to arrive at

\begin{equation}
 \label{expolaplinfini}
 [e^{i\xi L_n}]_{\vec{p}-\vec{q}} = [e^{i\xi L_n}]_{|p_1-q_1|,..,|p_n-q_n|} =
 \prod_{j=1}^n \frac{1}{(2\pi)}\int_{-\pi}^{\pi} e^{i(p_j-q_j)\kappa}e^{i2\xi (1-\cos{\kappa})}{\rm d}\kappa .
\end{equation}
Taking into account the definition of the modified Bessel functions of the first kind $I_p(z)= \frac{1}{\pi}\int_0^{\pi}e^{z\cos{\varphi}}\cos{p\varphi}{\rm d}\varphi$ where $p={\bf N}_0$
denotes non-negative integers \cite{abramo}, we can write the exponential matrix (\ref{expolaplinfini}) in the form

\begin{equation}
 \label{expobessel}
 [e^{i\xi L_n}]_{|p_1-q_1|,..,|p_n-q_n|} = e^{i 2n\xi} \prod_{j=1}^nI_{|p_j-q_j|}(-2i\xi) .
\end{equation}
Applying now the matrix relation (\ref{matrixpower}) and plugging in the exponential (\ref{expobessel}) yields an integral representation of the (negative semidefinite) fractional Laplacian matrix (\ref{fraalphn}) in terms of a product of
modified Bessel functions of the first kind, namely
\begin{equation}
 \label{fractionallapla}
 \begin{array}{l}
 \displaystyle [\Delta_{\alpha,n}]_{|p_1-q_1|,..,|p_n-q_n|}  = -\mu \Omega_{\alpha,n}^2[L_n^{\frac{\alpha}{2}}]_{|p_1-q_1|,..,|p_n-q_n|}   \\ \\
 \hspace{1cm} \displaystyle = -\mu \Omega_{\alpha,n}^2 \int_{-\infty}^{\infty} {\rm d}\xi \, e^{i 2n\xi} {\cal D}_{\frac{\alpha}{2}}(\xi) \prod_{j=1}^nI_{|p_j-q_j|}(-2i\xi) ,
 \end{array}
\end{equation}
with $-{\cal D}_{\frac{\alpha}{2}}(\xi)$ indicating the Riesz fractional derivative kernel of (\ref{distrifoufracder}).

\noindent {\bf Asymptotic behavior}\\
Introducing the new vector valued integration variable ${\vec{\xi}} = \vec{\kappa} p $ ($\xi_j=p\kappa_j , \forall j=1,..,n$) we can write for the infinite lattice integral of (\ref{spectralfrac})
by utilizing spherical polar coordinates $ \vec{p} =p \vec{e}_{\vec{p}} $  ($ \vec{e}_{\vec{p}}\cdot \vec{e}_{\vec{p}}=1$, $p^2=\sum_j^n p_j^2$)
\begin{equation}
 \label{recrit}
 L_n^{\frac{\alpha}{2}}({\bf p}) =  \frac{1}{(2\pi)^n} \int_{-\pi p}^{\pi p}..\int_{-\pi p}^{\pi p}\frac{{\rm d}\xi_1..{\rm d}\xi_n}{p^n}
 \left(4\sum_{j=1}^n\sin^2{\frac{\xi_j}{2p}} \right)^{\frac{\alpha}{2}}\cos{(\vec{\xi}\cdot \vec{e}_{\vec{p}})} .
\end{equation}
The dominating term for $p>>1$ becomes

\begin{equation}
 \label{recrit2}
 L_n^{\frac{\alpha}{2}}({\bf p}) \approx \frac{1}{p^{n+\alpha}} \frac{1}{(2\pi)^n} \int_{-\infty}^{\infty}..\int_{-\infty}^{\infty}{\rm d}\xi_1..{\rm d}\xi_n
\left( \sum_{j=1}^n \xi_j^2\right)^{\frac{\alpha}{2}}\cos{(\vec{\xi}\cdot \vec{e}_{\vec{p}} )} ,
\end{equation}
having the form
\begin{equation}
 \label{recrit3}
 L_n^{\frac{\alpha}{2}}(\vec{p})_{p>>1} \approx - \frac{C_{n,\alpha}}{p^{n+\alpha}} , \hspace{2cm} 0<\alpha< 2
\end{equation}
where the positive normalization constant is obtained explicitly as
$C_{n,\alpha}=\frac{2^{\alpha-1}\alpha\Gamma(\frac{\alpha+n}{2})}{\pi^{\frac{n}{2}}\Gamma(1-\frac{\alpha}{2})}$, e.g. \cite{michel-et-al-13,michel-et-al-14}, and holds only for 
$0<\alpha<2$. Relation (\ref{recrit3}) does not hold for $\alpha=2$ where the fractional Laplacian takes asymptotically for $\alpha\rightarrow 2-0$ 
the localized singular distributional representation $(-\Delta)\delta^n({\bf p})$ of the standard Laplacian.
The details of this limiting case calculation can be found in \cite{michel-et-al-14}.
We can identify the asymptotic representation (\ref{recrit2}), (\ref{recrit3}) with the kernel of Riesz fractional derivative (fractional Laplacian)
of the nD infinite space. For a more detailed discussion of properties we refer to \cite{michel-et-al-13,michel-et-al-14}.

\section{Conclusions}

In the present paper
we have developed a fractional lattice approach on nD periodic and infinite lattices. The fractional Laplacian matrices
conserve the `good' properties of the Laplacian matrices (translational symmetry and in our sign convention negative semi-definiteness).
The fractional lattice approach generalizes the concept of second order
centered difference operator appearing in the context of classical lattice models \cite{montroll}
to the concept of centered fractional order difference operator. For $\alpha=2$ the fractional
lattice approach contains the classical lattice approach, and for integer orders $\frac{\alpha}{2}\in \N$ finite centered differences of integer orders
of the second difference operator are generated.
For a discussion of properties of fractional Laplacian matrix on the cyclic chain, we refer to \cite{michelJphysA}.
In the infinite space and periodic lattice continuum limits these fractional Laplacian matrices take the representations of the well known respective Riesz fractional derivative kernels, i.e.
the convolutional kernels of the (continuous) fractional Laplacians. The approach allows to model `anomalous diffusion' phenomena on lattices with fractional transport phenomena including asymptotic emergence of L\'evy flights.
In such a fractional lattice diffusion model, the conventional Laplacian matrix is generalized by its fractional power law matrix function counterpart.
The formulation of our approach is consistent with recent works on the fractional approach developed on undirected networks by Riascos and
Mateos \cite{riascos-fracdyn,riascos-fracdiff}.
The present fractional lattice approach represents a point of departure to investigate anomalous diffusion and fractional random walk phenomena on lattices.
Such problems are defined by master equations
involving fractional Laplacian matrices such as deduced in the present work as generator matrices for the random dynamics. Fractional random walks on lattices and undirected networks
open currently a huge interdisciplinary research field \cite{michel-riascos,riascos-fracdyn,riascos-fracdiff}.
\section*{Acknowledgements}
{\it This work has been performed in
the framework of the ERCOFTAC SIG/42 cooperation during the sabbatical stay of F.C.G.A Nicolleau at
Institut Jean le Rond d’Alembert. We thank the two anonymous
reviewers for their useful remarks.}


\begin{thebibliography}{1}

\bibitem{abramo} Abramowitz and Stegun, Handbook of Mathematical Function, National Bureau of Standards: Applied Mathematics Series - 55, (1972), p. 376.
\bibitem{BornHuang} M. Born, K. Huang (1954) Dynamical Theory of Crystal Lattices, London: Oxford University Press.
\bibitem{gelfand} I.M. Gelfand, G.E. Shilov, Generalized Functions, Vol. I: Properties and Operations (New York:
Academic), (1964).
\bibitem{hilfer-2008} R. Hilfer, Threefold Introduction to Fractional Derivatives, in: Anomalous Transport: Foundations and Applications,
R. Klages et al. (eds.), Wiley-VCH, Weinheim, 2008, pp 17, ISBN: 978-3-527-40722-4.
\bibitem{Laskin} N. Laskin, Fractional Schr\"odinger equation, Phys. Rev. E 66, 056108 (2002).
\bibitem{laskin2006} N. Laskin, A. Zaslavsky, Nonlinear fractional dynamics on a lattice with long-range interactions, Physica A 368 (2006), 38-45.
\bibitem{metzler} R. Metzler, J. Klafter, The random walk's guide to anomalous diffusion: A fractional dynamics approach.
Physics Reports 339, pp. 1-77 (2000).
\bibitem{metzler2014} R. Metzler, J. Klafter, The restaurant at the end of the random walk: recent
developments in the description of anomalous transport by fractional dynamics, J. Phys. A: Math. Gen. 37 R161–R208 (2004).
\bibitem{michelJphysA} T.M. Michelitsch, B. Collet, A.F Nowakowski, F.C.G.A. Nicolleau,
Fractional Laplacian matrix on the finite periodic linear chain and its periodic Riesz fractional derivative continuum limit ,
J. Phys. A: Math. Theor. 48 295202  (2015). (arXiv:1412.5904).
\bibitem{michelchaos} T.M. Michelitsch, B. Collet, A.F. Nowakowski, F.C.G.A.
Nicolleau, Lattice fractional Laplacian and its continuum limit kernel on the finite cyclic chain, Chaos, Solitons \& Fractals 82, pp. 38-47 (2016).  	
(arXiv:1511.01251).
\bibitem{michel-collet} T. Michelitsch, B. Collet, X. Wang, Nonlocal constitutive laws generated by matrix functions:
Lattice dynamics models and their continuum limits, International Journal of Engineering Science {\bf 80}, 106–123 (2014).
\bibitem{michel-et-al-13} T.M. Michelitsch, G.A. Maugin, A.F. Nowakowski, F.C.G.A. Nicolleau, M. Rahman,
The Fractional Laplacian as a limiting case of a self-similar spring model and applications to n-dimensional
anomalous diffusion. Fractional Calclulus and Applied Analysis
vol. 16, no.4, 827-859 (2013).
\bibitem{michel-et-al-14} T.M. Michelitsch; G.A. Maugin, S. Derogar, M. Rahman, A regularized representation of the fractional Laplacian in n dimensions and its
relation to Weierstrass-Mandelbrot-type fractal functions, IMA Journal of Applied Mathematics 79, 753–777 (2014).
\bibitem{michel-riascos} T.M. Michelitsch, B.A. Collet, A.P. Riascos, A.F. Nowakowski, F.C.G.A Nicolleau, Fractional random walk lattice dynamics, to be published.
\bibitem{montroll} Maradudin, A. A., Montroll, E. A., \& Weiss, G. N. (1963). Theory of lattice dynamics in the harmonic approximation. Solid state of physics. New York: Academic
Press.
\bibitem{podlubny} I. Podlubny, Fractional Differential Equations, Mathematics in Science and Engineering, vol 198, Academic Press California 1999.
\bibitem{riascos12} A. P. Riascos, Jos\'e L. Mateos, Long-range navigation on complex networks using L\'evy random walks, Phys. Rev. E 86, 056110 (2012)
\bibitem{riascos-fracdyn} A.P. Riascos, J.L. Mateos, Fractional dynamics on networks:
Emergence of anomalous diffusion and L ́evy flights, Phys. Rev. E 90, 032809 (2014). (arxiv:1506.06167).
\bibitem{riascos-fracdiff} A.P. Riascos, J.L. Mateos, Fractional diffusion on circulant networks: emergence of a dynamical small world, J. Stat. Mech. (2015) P07015.
\bibitem{riascos15} A. P. Riascos and Jos\'e L. Mateos, Fractional quantum mechanics on networks: Long-range dynamics and quantum transport,
Phys. Rev. E 92, 052814 (2015).
\bibitem{riesz2} M.D. Ortiguera, Riesz Potential Operators and Inverses via Fractional Centered Derivatives,
International Journal of Mathematics and Mathematical Sciences, no. 48391, 1–12 (2006).
\bibitem{samko} S. Samko, A. Kilbas and O. Marichev, Fractional Integrals and Derivatives: Theory and Applications, Gordon and Breach, London (1993).
\bibitem{samko2003} S. Samko, Fractional Weyl-Riesz Integrodifferentiation of Periodic Functions of Two Variables
via the Periodization of the Riesz Kernel, Applicable Analysis, vol. 82, No 3, 269-299 (2003).
 \bibitem{taraII} V.E. Tarasov, Map of discrete system into continuous,
Journal of Mathematical Physics. Vol.47. No.9. (2006) 092901.
\bibitem{taraIII} V.E. Tarasov,
Continuous limit of discrete systems with long-range interaction,
Journal of Physics A. Vol.39. No.48. (2006) 14895-14910.  
\bibitem{tara1} V.E. Tarasov, Lattice fractional calculus, Applied Mathematics and Computation 257 (2015) 12–33.
\bibitem{tara2}  V.E. Tarasov, Exact discrete analogs of derivatives of integer orders: Differences as infinite series,
\bibitem{tara3} V.E. Tarasov, Exact discretization by Fourier transforms, Communications in Nonlinear Science and Numerical Simulation. Vol.37. (2016) 31-61.
Journal of Mathematics. Vol. 2015. (2015) Article ID 134842.
\bibitem{tara4} V.E. Tarasov, United lattice fractional integro-differentiation",
Fractional Calculus and Applied Analysis.
Vol.19. No.3. (2016) 625-664.
\bibitem{tara5} V.E. Tarasov, G.M. Zaslavsky, Fractional dynamics of coupled oscillators with long-range interaction, Chaos. Vol.16. No.2. (2006) 023110. 
\bibitem{Zoia2007} A. Zoia A, A. Rosso, M. Kardar, Fractional Laplacian in bounded domains,
Phys. Rev. E 76, 021116 (2007).

\end{thebibliography}
\end{document}